\documentclass{article}
\pdfoutput=1
\usepackage{spconf,amsmath,graphicx}
\usepackage{array,multirow}
\usepackage{url}
\usepackage{tikz}
\usepackage{subcaption}
\usepackage{ctable} 
\usepackage{xcolor}
\usepackage{textcomp}
\usepackage{hyperref}
\usepackage{lipsum}

\makeatletter
\def\thickhline{%
  \noalign{\ifnum0=`}\fi\hrule \@height \thickarrayrulewidth \futurelet
   \reserved@a\@xthickhline}
\def\@xthickhline{\ifx\reserved@a\thickhline
               \vskip\doublerulesep
               \vskip-\thickarrayrulewidth
             \fi
      \ifnum0=`{\fi}}
\makeatother

\newlength{\thickarrayrulewidth}
\newcommand{\mc}[2]{\multicolumn{#1}{#2}}
\usepackage{booktabs,siunitx}
\title{Low Resource audio-to-lyrics alignment from polyphonic music recordings}

\name{
Emir Demirel\textsuperscript{1}, Sven Ahlb\"ack\textsuperscript{2}, Simon Dixon\textsuperscript{1}
\thanks{ED received funding from the European Union's Horizon 2020 research and innovation programme under the Marie Skłodowska-Curie grant agreement No.\ 765068.}}

\address{
\textsuperscript{1}Centre for Digital Music, Queen Mary University of London, UK\\
\textsuperscript{2}Doremir Music Research AB, SE}

\ninept 
\begin{document}

\maketitle

\tikz [remember picture, overlay] %
\node [shift={(41mm,21.5mm)}] at (current page.south west) %
[anchor=south west] %
{\includegraphics[width=6mm]{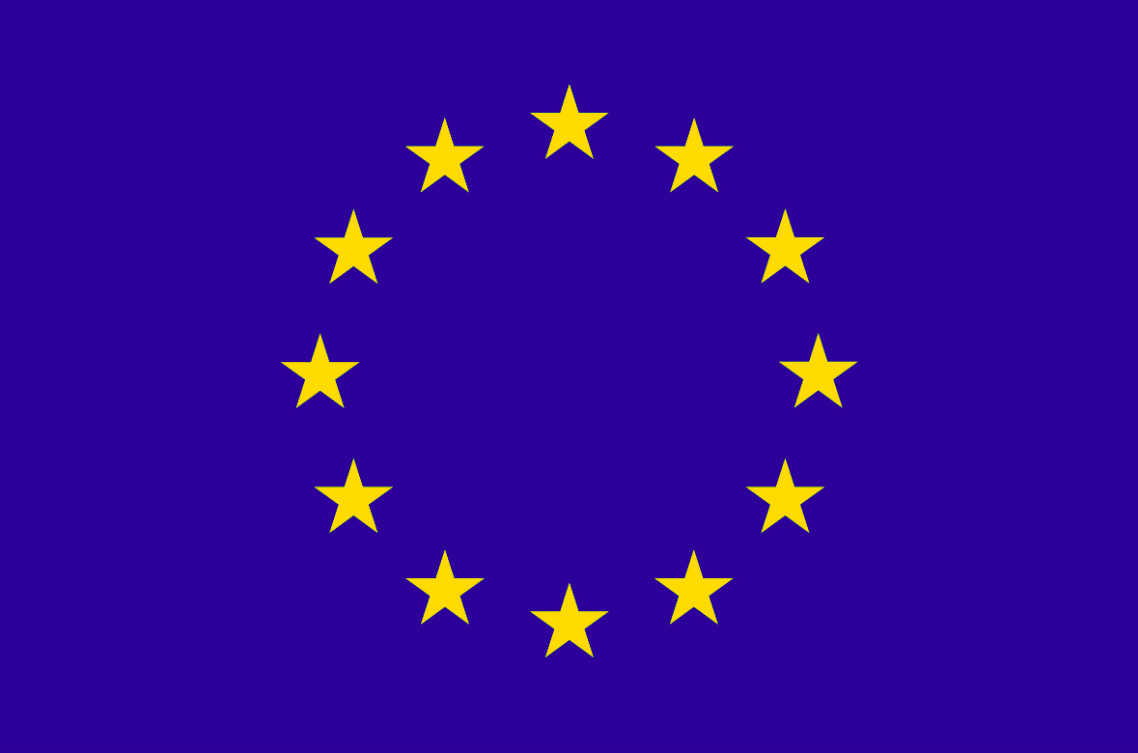}};

\begin{abstract}

Lyrics alignment in long music recordings can be memory exhaustive when performed in a single pass. In this study, we present a novel method that performs audio-to-lyrics alignment with a low memory consumption footprint regardless of the duration of the music recording. The proposed system first spots the anchoring words within the audio signal. With respect to these anchors, the recording is then segmented and a second-pass alignment is performed to obtain the word timings. We show that our audio-to-lyrics alignment system performs competitively with the state-of-the-art, while requiring much less computational resources. In addition, we utilise our lyrics alignment system to segment the music recordings into sentence-level chunks. Notably on the segmented recordings, we report the lyrics transcription scores on a number of benchmark test sets. Finally, our experiments highlight the importance of the source separation step for good performance on the transcription and alignment tasks. For reproducibility, we publicly share our code with the research community.
\end{abstract}
\section{Introduction}
\label{sec:intro}

Audio-to-lyrics alignment has a variety of applications within the music technology industry. Some of these applications include lyrics prompting for karaoke applications, captioning, and music score and video editing. Moreover, lyrics alignment systems can be leveraged to generate new training data for several tasks within MIR research, such as lyrics transcription, singing voice detection, source separation, music transcription and cover song identification.

The task of aligning song lyrics with their corresponding music recordings is among the most challenging tasks in music information retrieval (MIR) research due to three major factors: the multi-modality of the information to be processed -- namely \textit{music} and \textit{speech}, the presence of the musical accompaniment in the acoustic scene and the length of the music recording to be aligned. For processing linguistically relevant information, previous studies have taken the approach of adapting automatic speech recognition (ASR) paradigms to singing voice signals \cite{dzhambazov2015modeling, stoller2019end,gupta2019automatic,sharma2019automatic}. Regarding the musical accompaniment, researchers have aligned lyrics on either source separated vocal tracks \cite{sharma2019automatic} or utilized acoustic models trained on polyphonic recordings \cite{stoller2019end,gupta2019automatic,vaglio2020}.

For relevant alignment tasks within MIR research, previous studies have presented efficient ways to handle the duration issue in alignment by using methods based on dynamic time warping (DTW) \cite{muller2004towards,dixon2005line}. The results in the MIREX 2017\footnote{Can be accessed from \url{https://www.music-ir.org/mirex/wiki/2017:Automatic_Lyrics-to-Audio_Alignment_Results}} public evaluation of audio-to-lyrics alignment systems showed that Viterbi-based alignment outperforms DTW \cite{kruspelyrics}. Most lyrics alignment research since then has utilized a similar approach due to its performance and efficiency, however even Viterbi decoding may become intractable when processing long audio recordings. Within this context, performing robust and efficient lyrics alignment in long music recordings still remains as a bottleneck and preventing factor for audio-to-lyrics alignment technology in industrial applications involving large-scale web services or mobile deployment. In this paper, we aim to contribute to this field of research by proposing a low resource solution that is robust and efficient in terms of the audio length and the musical accompaniment to singing. Leveraging its duration-invariant capability, we further show that this approach could be exploited to generate sentence-level lyrics annotations for extending existing lyrics transcription training sets.

This paper continues as follows: in Section \ref{sec:related}, we provide a brief overview of the state-of-the-art in audio-to-lyrics alignment. We explain the overall details of the proposed biased lyrics search pipeline in Section \ref{sec:system}. Then, we evaluate the utilization capability of the overall system through lyrics alignment and transcription experiments. Also in this section, we conduct the first experiments evaluating different source separation algorithms in the lyrics alignment and transcription context. Finally, we report results that are competitive with the state-of-the-art in lyrics alignment and best recognition scores for lyrics transcription on a public benchmark evaluation dataset.

\section{Related Work}
\label{sec:related}

Early audio-to-lyrics alignment systems in research use Hidden Markov Model (HMM) based acoustic models which are utilized to extract frame-level phoneme (or character) posterior probabilities. Then a forward-pass decoding algorithm is applied on these posteriograms, obtaining phoneme alignments. Then using a language model (LM), phoneme posteriograms can be converted to word posteriograms to retrieve word-level alignments \cite{dzhambazov2015modeling,gupta2019automatic,sharma2019automatic}. One recent successful system \cite{stoller2019end} showed a considerable performance boost compared to previous research using an end-to-end approach trained on a large corpus, where alphabetic characters are used as sub-word units of speech. Vaglio et al.\ \cite{vaglio2020} also employed an end-to-end approach for lyrics alignment applied in a multilingual context, but using phonemes as an additional intermediate representation, and obtained competitive results. In addition, the authors have used a public dataset \cite{meseguer2019dali} that is much smaller than the training set used in \cite{stoller2019end}. Gupta et al.\ \cite{gupta2019automatic} reported state-of-the-art results using an acoustic model trained on polyphonic music using genre-specific phonemes. According to the authors, their system applies forced alignment with a large beam size as their system attempts to process the entire music recording at once. 
Although achieving impressive results, the application of forced alignment in a single pass can be memory exhaustive.

A similar challenge within automatic speech recognition (ASR) research is the \textit{lightly supervised alignment} task \cite{bell2015mgb} where the goal is to obtain timings of human-generated captions in long TV broadcasts that would be displayed to TV viewers as subtitles. Moreno et al.\  \cite{moreno1998recursive} present a system for long audio alignment which searches for the input word locations through a recursive application of speech recognition on a gradually restricted search space and language model. The method is then further improved in terms of robustness and efficiency in the search space using the factor transducer approach \cite{bell2015system,moreno2009factor}. The factor transducer introduces an important drawback within the lyrics alignment task as it exerts weak timing constraints during decoding, i.e. the output word alignments are not constrained to be in order. This would rise as a significant issue especially during successive patterns of similar word sequences, which occur frequently in song lyrics \cite{demirex2020mirex}.
 
Another major challenge during lyrics alignment (and also transcription) is the influence of accompanying non-vocal musical sounds.  One way to minimize this during the transcription and alignment is by isolating the vocal track using a vocal source separation system. There has been a number of powerful open-source music source separation toolkits made available for research recently \cite{stoter2019open,hennequin2019spleeter,defossez2019music}, yet the output vocal track is not guaranteed to be free of sonic artifacts introduced during separation. In turn, these artifacts might affect the word intelligibility and thus the accuracy during the automatic lyrics transcription and alignment (ALTA) tasks. The effects of different source separation algorithms on lyrics transcription and alignment has not been studied extensively; we provide a comparison in this paper.

\section{System Details}
\label{sec:system}

\begin{figure}[ht]
 \centering
 \includegraphics[clip,width=0.45\textwidth,height=6cm]{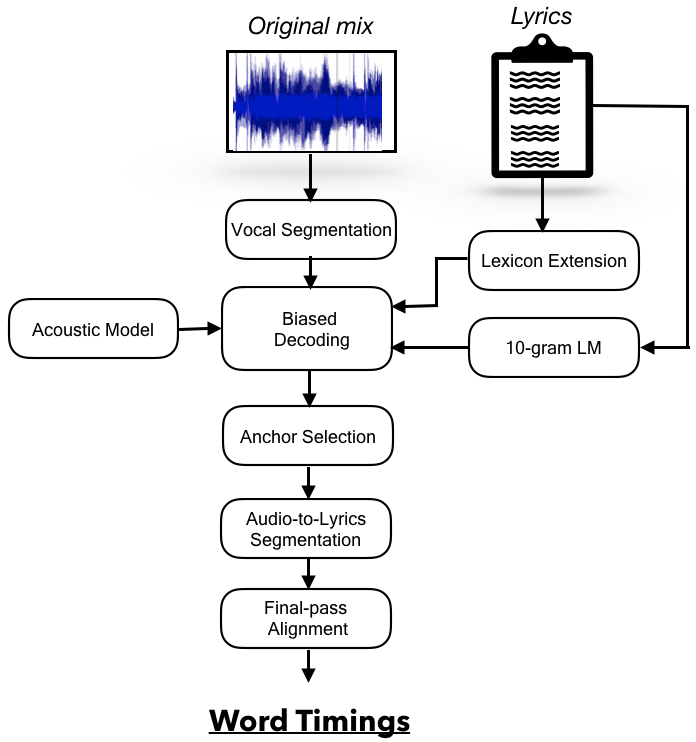}% 
 \caption{Pipeline of our lyrics transcription and alignment system}
 \label{fig:block_diag}
\end{figure}

Our overall lyrics alignment pipeline is illustrated in Figure \ref{fig:block_diag}. We initially extract vocals from the original polyphonic mix and retrieve vocal segments using energy-based voice activity detection. We search for lyrics within these segments by applying automatic transcription using a decoding graph constructed via a biased language model (LM). The matching portions of the transcribed and original lyrics and their location in the audio signal are obtained through the text and forced alignment techniques respectively. The music recording is then segmented with respect to these matching portions and a final-pass alignment is applied to obtain the timings of all the words in the original lyrics. In order to be able to align and recognize out-of-vocabulary (OOV) words during decoding, we extend the pronunciation dictionary for the words in the input lyrics.

\subsection{Extending the Pronunciation Dictionary}

For achieving a robust lyrics alignment system, out-of-vocabulary (OOV) words have to be taken into account. Lyrics may contain linguistically ambiguous sequences of words such as `la', `na' and `ooh', as well as words with repeated syllables or phonemes like `do re mi fa so oh oh oh', out-of-language (OOL) words or special names. Thus, we extend the pronunciation dictionary with respect to the input lyrics, and construct a pronunciation transducer prior to decoding.
We trained a grapheme-to-phoneme conversion (\textit{g2p}) model \cite{novak2011phonetisaurus}, using the CMU English pronunciation dictionary\footnote{\url{http://www.speech.cs.cmu.edu/cgi-bin/cmudict}}, to generate new pronunciations for each OOV word in the input lyrics.

\subsection{Vocal Segmentation}

Initially, we separate the vocal track from the original mix and determine the voice activity regions (VAR) based on the log-energy (the zeroth component of MFCC features). We merge consecutive VARs if the silence between them is less than $\tau_\mathit{silence}$ seconds, although we do not merge segments that are already more than $\tau_\mathit{max}$ seconds long. The values for $\tau$ are determined empirically. Note that if $\tau_\mathit{silence}$ is too short, there occurs the risk of over-segmenting the audio which could potentially reduce the word recognition rate. We have set $\tau_\mathit{silence} = 0.8$ and $\tau_\mathit{max} = 6$, based on our empirical observations.

\subsection{Biased Decoding}

The goal of this stage is to detect the word positions in the lyrics and the audio that we will use for segmentation later on. In order to be able to detect the words in input lyrics with a higher recognition rate, we restrict the search space by building an n-gram language model (LM) on the input lyrics. First, we transcribe the contents within the VARs using the pretrained acoustic model in \cite{demirel2020} and the biased LM. Song lyrics often contain repetitive word sequences for which the system might recognize a word in the wrong position or repetition in the lyrics, potentially causing accumulated errors in segmentation. For robustness against such cases, we exert strong constraints on the word order via building the LM with 20-grams. Moreover, prior to processing, we add a $<$\textit{NOISE}$>$ tag at the beginnings and endings of each lyrics line to further reduce the risk of alignment errors between long non-vocal segments.

Building the LM from only the input lyrics has two major advantages: First, it increases the chance of recognizing the words in input lyrics while minimizing the risk of wrong word predictions. Secondly, through constructing the LM on the fly, we do not need an external pretrained LM to perform lyrics alignment. This aspect of our system makes it applicable in a multilingual setting in the presence of a g2p model for target languages.

\subsection{Anchor Selection}

\begin{figure}[ht]
 \centering
 \includegraphics[clip,width=0.42\textwidth,height=3.2cm]{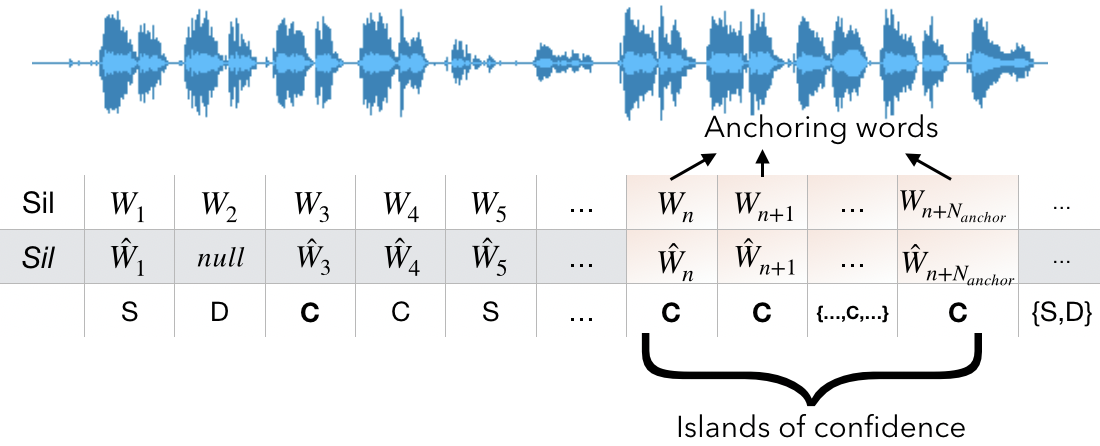}% 
 \caption{Anchor selection. $W_n$ and $\hat{W}_n$ are the reference and predicted words respectively. D and S stand for word deletions and substitutions after text alignment. C are the labels for correctly recognized (matching) words. }
 \label{fig:anchor}
\end{figure}

Next, we apply text alignment between the transcriptions and the reference lyrics to determine the matching portions, i.e.\  \textit{anchoring} words. To impose further constraints on word order, we consider $N_\mathit{anchor}$ number of successive matching words as the anchoring instances between the lyrics and the audio signal. On the corresponding VARs, we apply forced alignment using these anchoring words to get their positions on the audio signal. We refer these portions of the audio signal matched with its lyrics as \textit{islands of confidence}.  Using a large $N_\mathit{anchor}$ would carry the risk of detecting lesser anchoring words while a very small value could cause over-segmentation. In our experiments, we chose  $N_\mathit{anchor}=5$ (Fig.\  \ref{fig:anchor}). 

\subsection{Audio and Lyrics Segmentation}\label{sec:seg}

Once detected, the anchoring words are utilized to split the music recording into shorter segments. In order to further alleviate the risk of over-segmenting the recording, we allow a maximum of $N_\mathit{segment}$ anchoring words to be in each segment. We start segmenting monotonically from the beginning to the very end of the recording. Once $N_\mathit{segment}$ words are spotted, the audio is split with respect to the beginning of the first and the ending of the $N_\mathit{segment}$-th word. Empirically, we found this approach to function well for $N_\mathit{segment} > 10$. If the first word in the original lyrics is not spotted, the beginning of the first segment in the initial voice-activity-based segmentation is used. A similar approach is applied for the last word.

\subsection{Final-Pass Alignment}

Finally, we can apply forced alignment on shorter audio segments which are extracted in the previous step using a smaller beam size. In our experiments, we were able to obtain alignments without any memory issues using a beam size of 30 and a retry beam size of 300, which are much lower than the values reported in \cite{sharma2019automatic}.

\section{Experimental Setup}
\label{sec:illust}

 To test the quality of the output audio-to-lyrics segmentation, we evaluate the precision of the word timings produced by the final pass alignment. Further, we evaluate the transcription performance on the initial voice-activity segments and the final segmentation, to gain an insight into whether these segments can be used for further training.

\textit{Vocal Extraction : }
The alignment and decoding are applied on the separated vocal tracks, in order to minimize the influence of accompanying musical sounds. Additionally, the acoustic model employed in decoding is trained on monophonic singing voice recordings \cite{demirel2020}. The output of the vocal source separation has a direct effect on the performance of decoding, and hence the overall performance of lyrics alignment. There are two mainstream approaches in vocal separation: spectrogram-based and waveform-based models. While spectrogram-based approaches have been more widely used \cite{stoller2018wave,stoter2019open}, there have been recent successful waveform-based music source separation methods, motivated by capturing the phase information in the signal \cite{hennequin2019spleeter}. To test the effect of the source separation step, we compare a waveform-based model (Demucs) \cite{defossez2019music} and a state-of-the-art spectrogram-based model (Spleeter) \cite{hennequin2019spleeter} in experiments.

\textit{Lyrics Transcriber : }
We use the acoustic model of the lyrics transcriber in \cite{demirel2020}. The system was trained on 150 hours of a cappella singing recordings with a non-negligible amount of noise \cite{smule}.  After decoding with a 4-gram LM, we rescore lattices with RNNLM \cite{xu2018pruned} and report this value as the final transcription result.

\textit{Data : }
For testing, we use the benchmark evaluation set for the  lyrics transcription and alignment tasks, namely JamendoLyrics \cite{stoller2019end} which consists of 20 music recordings (1.1 hours) from a variety of genres including metal, pop, reggae, country, hiphop, R\&B and electronic. In addition, the lyrics transcription results are reported also on the Mauch \cite{mauch2011integrating} dataset which consists of 20 English language pop songs.

\section{Results \& Discussion}

\subsection{Audio-to-Lyrics Alignment}

The word alignments are evaluated in terms of the standard audio-to-lyrics alignment metrics: mean and median average absolute error (AE) \cite{mesaros2008automatic} in seconds, and correctly predicted segments (PCS) with a tolerance window of 0.3 seconds \cite{mauch2011integrating}. These metrics are computed for each sample in the evaluation set and the mean for each metric is reported as the final result.

We compare the performance of our system with the most recent successful lyrics alignment systems. SD1 \cite{stoller2019end} applies alignment on polyphonic recordings using an end-to-end system trained on a private dataset consisting of over 44\,000 songs. In SD2, alignment is performed on source separated vocal track using the Wave-U-Net \cite{stoller2018wave} architecture. VA \cite{vaglio2020} also uses an end-to-end model, but trained on the DALI (v1.0) dataset, which has over 200 hours of polyphonic music recordings and extracts the vocals using Spleeter. GC1 \cite{gupta2019automatic} uses the same training data, for constructing an acoustic model in the hybrid-ASR setting \cite{povey2016purely} and performs alignment on the original polyphonic mix as well. In addition to these models, we refer to our models which align words to the vocal tracks separated by Demucs and Spleeter as DE1 and DE2 respectively. 

\begin{table}[ht]
  \centering
  \scalebox{0.9}{
  \begin{tabular}{l r | r | r}
     & {Mean AE}   & {Median AE} & {PCS} \\ \thickhline
     SD1 \cite{stoller2019end}   & 0.82   &  0.10 &  0.85  \\ \hline
     SD2 \cite{stoller2019end}  & 0.39   &  0.10 &  0.87 \\ \hline
     VA \cite{vaglio2020}  & 0.37   &  N/A &  0.92 \\ \hline     
     GC1 \cite{gupta2019automatic}  & \textbf{0.22}   & \textbf{0.05}  &  \textbf{0.94}  \\ \thickhline
     DE1  &    0.31    & \textbf{0.05}  & 0.93 \\ \hline
     DE2 & 0.38  &  \textbf{0.05} &  0.90 \\ 
  \end{tabular}}
  \caption{Lyrics alignment results on the Jamendo dataset}
  \label{table:a2l}
\end{table} 

According to Table \ref{table:a2l}, our system that uses the waveform-based source separation model (Demucs) for vocal extraction outperforms other methods that use end-to-end learning, and obtains competitive results to the state-of-the-art (GC1). Using Spleeter, we were able to achieve similar results to VA \cite{vaglio2020}, which also uses Spleeter for source separation. Note that the training data we have used is smaller and less diverse than the datasets used by all other systems, highlighting that there is room for performance improvement in our method. Moreover, the alignment performance is better when using Demucs instead of Spleeter in terms of mean AE and PCS, even though the median AE is the same.

\begin{figure}[!ht]
 \centering
 \includegraphics[clip,width=0.3\textwidth,height=1.7cm]{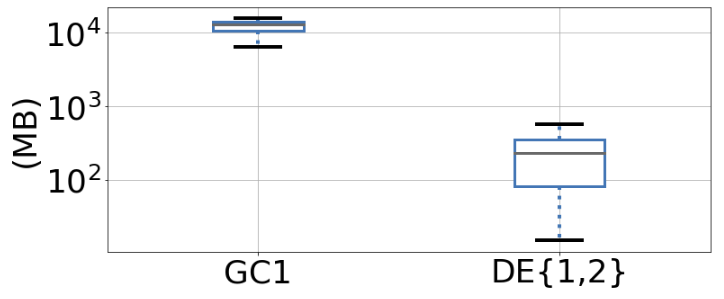}% 
 \caption{Memory usage on RAM in megabytes (MB)}
 \label{fig:memory}
\end{figure}

Figure \ref{fig:memory} shows a comparison of the system presented in this paper, DL\{1,2\}, and GL1 in regard to memory efficiency. The memory usage on the RAM is monitored every 10 seconds during a run over the Jamendo dataset. We have executed the code for the experiments on Intel$^{\tiny{\textrm{\textregistered}}}$ Xeo\textsuperscript{TM}  E5-2620 with 24 GM of RAM capacity. Below, Figure \ref{fig:memory} and Table \ref{table:mem} show that the memory consumption of our system is less than GL1 by a margin more than an order of magnitude. The larger standard deviation in our case is potentially due to the varying lengths and complexities of the segmented music signals input to decoding.

\begin{table}[!h]
\centering
\scalebox{0.9}{
\begin{tabular}{ l | l  l}
& {GC1} &  {DE\{1,2\}} \\  \hline
Mean (Std.\%) & 13,740 (8.8\%) & \textbf{343} (31\%)\\\hline
Max & 16,745 &  \textbf{748} \\\hline
\end{tabular}}
\caption{Statistics on memory usage in MB}
 \label{table:mem}
\end{table}

\subsection{Automatic Lyrics Transcription}

In order to gain an insight as to whether the final lyrics-to-audio segments can be leveraged as sentence-level annotations for training data, we conduct lyrics transcription experiments. We compare word recognition rates for a pure inference case on VARs and on the segments extracted as described in Section \ref{sec:seg}. Additionally, we provide comparisons with the state-of-the-art in lyrics transcription from polyphonic music recordings: SD1 \cite{stoller2019end}, GC1 and GC2 \cite{gupta2019automatic} (vocals extracted using \cite{stoller2018wave}). For the evaluation, we use word (WER) and character (CER) error rate computed using the Kaldi toolkit. 

According to Table \ref{table:alt}, unlike the alignment results, using Spleeter for separating the vocals seems to be more beneficial for lyrics transcription. Notice that while the gap is small between the recognition rates of DE1 and DE2 on the Mauch dataset, it is much higher on the Jamendo dataset. For both DE1 and DE2, the transcription rates are consistently higher on the audio-to-lyrics segments compared to VAR, implying that our segmentation method can be exploited in a semi-supervised setting. 

\begin{table}[ht]
\centering
\scalebox{0.86}{
\begin{tabular}{ l r r | r r r}
& \mc{2}{c}{\textit{WER}}    
 & \mc{2}{c}{\textit{CER}}  \\
\cmidrule{2-5}
& {Mauch} &  {Jamendo}  
& {Mauch} &  {Jamendo}    \\
\cmidrule{2-5} 
SD1 \cite{stoller2019end} &   70.09    &  77.80  &    48.90    &  49.20   \\ \cmidrule{1-5}
GC1 \cite{gupta2019automatic} &  \textbf{\textit{44.02}}    & \textit{ 59.57} & N/A & N/A  \\ \cmidrule{1-5}
GC2 \cite{gupta2019automatic} &  78.85   &  71.83 & N/A & N/A  \\ \cmidrule{1-5}
DE1 - VAR  &   60.92    &   62.55 &  44.15    &   47.02 \\ \cmidrule{1-5}
DE1 - segmented  &  50.44 &  55.47  &  38.65 &  42.11 \\ \cmidrule{1-5}
 DE2 - VAR  &   57.36    &  \textbf{51.76} &   \textbf{41.52}    &  \textbf{37.26} \\ \cmidrule{1-5}
 DE2 - segmented & 49.92 &  \textbf{44.52} &   \textbf{38.41}  &  \textbf{32.90} & \\ 
\end{tabular}}
\caption{Lyrics transcription results}
 \label{table:alt}
\end{table}

During pure inference on VAR, our system outperforms the state-of-the-art on Jamendo, while the performance on the Mauch dataset is still far behind. In all cases, our system outperforms other systems that apply transcription on separated vocal tracks by a great margin. Note this could be due to the fact that all of the previous methods use datasets consisting of polyphonic recordings even though they are much larger in size.

Overall, the results show that the application of lyrics transcription and alignment on separated vocal tracks can achieve competitive performance to state-of-the-art systems working directly on polyphonic music input. It should be noted that the choice of source separation has a crucial but inconsistent effect on the final transcription and alignment results. While it is not yet possible to draw a general conclusion regarding which method works better for which task, we have shown that the effect of vocal extraction on the performance of these tasks is worthy to consider when developing music source separation algorithms.

\section{Conclusion}

We presented a novel system that segments polyphonic music recordings with respect to its given lyrics, and further aligns the words with the audio signal. We have reported competitive results with the state-of-the-art while outperforming other end-to-end based models.  Through lyrics transcription experiments, we provided evidence supporting the capability of our system to be exploited for generating additional training data for a variety of MIR tasks. As a pilot study, we have conducted the first experiments on the effect of different source separation models on the lyrics transcription and alignment tasks. The recognition rates on separated vocals show that our system performs better than the previous best systems by a considerable margin while showing comparable performance with the state-of-the-art in ALT from polyphonic music recordings on a public benchmark evaluation set. Moreover, it is shown that our acoustic model can be exploited for both of the ALTA tasks. 

It should be noted that the system presented here achieves this performance requiring considerably lower computational and data resources compared to the best performing published work to this date, which is shown via a quantitative comparison regarding the memory consumption during runtime. From this perspective, this framework is shown to be applicable in use cases where low resource solutions are required, such as large-scale web services and mobile applications. As an additional advantage, our approach does not rely on a pretrained language model, which makes it possible for it to be extended for a multilingual setup. As a final remark, we have publicly shared the code\footnote{Can be accessed from \url{https://github.com/emirdemirel/ASA_ICASSP2021}} for open science and reproducibility.

\begin{tikzpicture}[remember picture,overlay]
\node[anchor=south,yshift=10pt] at (current page.south)
{\fbox{\parbox{\dimexpr\textwidth-\fboxsep-\fboxrule\relax}{
\footnotesize  Copyright 2021 IEEE. Published in the IEEE 2021 International Conference on Acoustics, Speech, and Signal Processing (ICASSP 2021), scheduled for 6-11 June, 2021, in Toronto, Canada. Personal use of this material is permitted. However, permission to reprint/republish this material for advertising or promotional purposes or for creating new collective works for resale or redistribution to servers or lists, or to reuse any copyrighted component of this work in other works, must be obtained from the IEEE. Contact: Manager, Copyrights and Permissions / IEEE Service Center / 445 Hoes Lane / P.O. Box 1331 / Piscataway, NJ 08855-1331, USA. Telephone: + Intl. 908-562-3966.
}}};
\end{tikzpicture}

\bibliographystyle{IEEEbib}
\bibliography{strings,refs}

\end{document}